\documentclass{Interspeech2024}
\pdfoutput=1

\interspeechcameraready

\usepackage[acronym, shortcuts, nohypertypes={acronym}]{glossaries}
\usepackage{color}
\newacronym{AI}{AI}{Artificial Intelligence}
\newacronym{AST}{AST}{Audio Spectrogram Transformer}
\newacronym{ASR}{ASR}{Automatic Speech Recognition}
\newacronym{BERT}{BERT}{Bidirectional Encoder Representations from Transformers}
\newacronym{BPTT}{BPTT}{Backpropagation Through Time}
\newacronym{ComParE}{ComParE}{Computational Paralinguistics Challenge}
\newacronym{CNN}{CNN}{Convolutional Neural Network}
\newacronym{CRF}{CRF}{Conditional Random Field}
\newacronym{defcomm}{\textsc{DefComm-DB}}{Multimodal Defensive Communication Database}
\newacronym{DMRS}{DMRS}{Defense Mechanisms Rating Scale}
\newacronym{DMRS-Q}{DMRS-Q}{Q-sort based Defense Mechanisms Rating Scale}
\newacronym{egemaps}{eGeMAPS}{extended Geneva Minimalistic Acoustic Parameter Set}
\newacronym{ELECTRA}{ELECTRA}{Efficiently Learning an Encoder that Classifies Token Replacements Accurately}
\newacronym{FAU}{FAU}{Facial Action Unit}
\newacronym{GRU}{GRU}{Gated Recurrent Unit}
\newacronym{HNR}{HNR}{Harmonics to Noise Ratio}
\newacronym{LLD}{LLD}{Low-Level Descriptor}
\newacronym{LM}{LM}{Language Model}
\newacronym{LLM}{LLM}{Large Language Model}
\newacronym{LPC}{LPC}{Linear Predictive Coding}
\newacronym{LSP}{LSP}{Line Spectral Pair}
\newacronym{LSTM}{LSTM}{Long Short-Term Memory}
\newacronym{MFA}{MFA}{Montreal Forced Aligner}
\newacronym{MFCC}{MFCC}{Mel-Frequency Cepstral Coefficient}
\newacronym{ML}{ML}{Machine Learning}
\newacronym{MLM}{MLM}{Masked Language Model}
\newacronym{openSMILE}{openSMILE}{Open-source Speech and Music Interpretation by Large-space Extraction}
\newacronym{OpenCV}{OpenCV}{Open Source Computer Vision}
\newacronym{PLP-CC}{PLP-CC}{Perceptual Linear Prediction and Cross-Correlation}
\newacronym{Py-Feat}{Py-Feat}{Python Facial Expression Analysis Toolbox}
\newacronym{RBF}{RBF}{Radial Basis Function}
\newacronym{RNN}{RNN}{Recurrent Neural Network}
\newacronym{SER}{SER}{Speech Emotion Recognition}
\newacronym{SSD}{SSD}{Spectral Shape Descriptor}
\newacronym{SMOTE}{SMOTE}{Synthetic Minority Over-sampling Technique}
\newacronym{SVM}{SVM}{Support Vector Machine}
\newacronym{UAR}{UAR}{Unweighted Average Recall}
\newacronym{ViT}{ViT}{Vision Transformer}
\newacronym{W2V}{W2V}{Wav2Vec 2.0}
\newacronym{WER}{WER}{Word Error Rate}

\newcommand{\ie}{i.\,e., }
\newcommand{\eg}{e.\,g., }

\usepackage{hyperref}
\usepackage{cleveref}

\title{This Paper Had the Smartest Reviewers -- \\ Flattery Detection Utilising an Audio-Textual Transformer-Based Approach}
\name[affiliation={1}]{Lukas}{Christ}
\name[affiliation={2}]{Shahin}{Amiriparian}
\name[affiliation={3}]{Friederike}{Hawighorst}
\name[affiliation={3}]{Ann-Kathrin}{Schill}
\name[affiliation={3}]{Angelo}{Boutalikakis}
\name[affiliation={4}]{Lorenz}{Graf-Vlachy}
\name[affiliation={3}]{Andreas}{König}
\name[affiliation={1,2,5}]{Björn W}{Schuller}

\address{
  $^1$University of Augsburg, Germany 
  $^2$CHI, TU Munich, Germany
  $^3$University of Passau, Germany\\
  $^4$TU Dortmund, Germany
  $^5$GLAM, Imperial College London, UK
  }
\email{lukas1.christ@uni-a.de}

\usepackage[activate]{microtype}
\begin{document}

\maketitle
 
\begin{abstract}

Flattery is an important aspect of human communication that facilitates social bonding, shapes perceptions, and influences behavior through strategic compliments and praise, leveraging the power of speech to 
build rapport effectively.
Its automatic detection can thus enhance the naturalness of human-AI interactions. To meet this need, we present a novel audio textual dataset comprising $20$ hours of speech and train machine learning models for automatic flattery detection. In particular, we employ pretrained AST, Wav2Vec2, and Whisper models for the speech modality, and Whisper TTS models combined with a RoBERTa text classifier for the textual modality. Subsequently, we build a multimodal classifier by combining text and audio representations. Evaluation on unseen test data demonstrates promising results, with Unweighted Average Recall scores reaching $82.46$\% in audio-only experiments, $85.97$\,\% in text-only experiments, and $87.16$\,\% using a multimodal approach.
\end{abstract}

\section{Introduction}

Flattery is a pervasive social influencing behavior in which one individual (i.e., the flatterer) provides deliberate appreciation toward another individual or group (i.e., the flattered)~\cite{jones1975ingratiation}. Specifically, flattery accentuates the positive qualities of the flattered with the objective of interpersonal attractiveness and ultimately winning favour~\cite{kumar1991construction, westphal1998board}. While flattery is rather difficult to detect for the flattered, it is easier to be detected by bystanders~\cite{vonk2007ingratiation}. As such, flattery is one of the most commonplace social influencing behaviors that many individuals employ, receive, or witness on a daily basis~\cite{gordon1996impact, vonk2007ingratiation}. 
Research across various disciplines has extensively explored the potency of flattery, particularly within organizational contexts. Kumar and Beyerlein~\cite{kumar1991construction}, for instance, devised the Measure of Ingratiatory Behaviors in Organizational Settings (MIBOS) scale to unveil how employees adeptly employ flattery as an upward influencing technique towards immediate superiors~\cite{kumar1991construction}. Likewise,~\cite{ellis2002use} demonstrated that job interviewees may employ flattery toward their job interviewers to enhance these job interviewers’ evaluations. In the broader management context, it was shown that journalists exhibit more positive coverage of a firm and its CEO when flattered by the CEO and capital markets analysts also respond to CEO flattery by issuing more positive firm ratings~\cite{westphal2008sociopolitical, westphal2011avoiding}.

Motivated by the relevance of flattery in interpersonal communication, we investigate its automatic detection via \ac{ML} methods, utilizing speech data. Such a methodology may be applicable in, \eg human-computer interaction, communication training~\cite{mcewen1997communication}, and computational psychometrics~\cite{cipresso2017back}. Automatic analysis of speech for understanding different facets of human communication is an active research field, with \ac{SER} arguably being the most prominent task (\eg~\cite{gerczuk2021emonet, pepino21_interspeech,wagner2023dawn}). Other examples include the detection of humour (\eg~\cite{christ2023towards, amiriparian2022muse,hasan2019ur}), sarcasm~\cite{bedi2021multi}, interpersonal adaptation~\cite{amiriparian2019synchronization}, and defensive communication~\cite{defensive2023}. Oftentimes, it is beneficial to also consider linguistic information, \ie transcripts of the speech samples, when addressing such tasks~\cite{amiriparian2021impact,gandhi2023multimodal, christ2023muse}. 
To the best of our knowledge, our work is the first to consider the task of detecting flattery from speech. Our contributions are as follows: (i) we introduce a novel dataset for flattery detection from speech; (ii) we provide \ac{ML} approaches for automatic flattery detection.

\section{Dataset}
Considering the large body of literature on flattery in the management context, we analyze flattery as applied by business analysts. We obtain a dataset of $2159$ dyads of analyst questions and CEO answers within earnings calls of large, publicly listed U.\,S.\ hard- and software as well as pharmaceutical firms from 2013 to 2018, which were transcribed and published by Seeking Alpha.
Drawing from prior research~\cite{konig2018silver, krippendorff2018content}, we systematically develop a reliable, context-sensitive, content-analytical measure for detecting and assessing analyst flattery. A detailed account of our annotation guidelines is provided in~\cite{schill2022analyst}. A team of three expert human annotators label the dataset for instances of flattery on a span-level, \ie flattery is identified in subsentential word sequences. An example is considered flattery if and only if all three annotators agree on it.

    In our machine learning experiments, the problem of flattery detection is framed as a sentence-level binary classification task here. The transcripts are split into sentences via the pySBD~\cite{sadvilkar-neumann-2020-pysbd} tool. We then project the subsentential annotations to the sentence level by treating a sentence as a positive example of flattery if it contains a passage identified as flattery by the annotators. To build the audio dataset, we first reconstruct word-level timestamps from the speech data using the \ac{MFA}~\cite{mcauliffe2017montreal} library. Based on these timestamps, the speech recordings are cut such that each resulting audio sample corresponds to one sentence. In total, 10,903 such samples uttered by $255$ different speakers are obtained, of which $752$ ($6.90\,\%$) are considered flattery. Overall, the dataset consists of almost $20$ hours of speech, with an average sample length of $6.59$ seconds.

    A speaker-independent splitting into a training, development, and test set is created, where the training partition comprises $70\,\%$ of the speakers ($178$/$255$) while $15\,\%$ of the speakers are assigned to the development and test partition each. We ensure that the fraction of positive examples as well as the mean duration of samples in each partition is comparable. A detailed overview of the resulting dataset is provided in~\Cref{tab:stats}.


\begin{table}[th]
  \caption{Dataset statistics.}
  \label{tab:stats}
  \centering\resizebox{1\columnwidth}{!}{
  \begin{tabular}{lrrrr}
    \toprule
     & \multicolumn{1}{r}{train} & \multicolumn{1}{r}{development} & \multicolumn{1}{r}{test} & \multicolumn{1}{r}{total}\\
    \midrule
    \#\,speakers (m, f) & 178 (162, 16) & 39 (35, 4) & 38 (35, 3) & 255 (232, 23) \\
    \#\,samples (flattery) & 7167 (6.7\%) & 1878 (7.4\%)‌ & 1858 (7.2\%) & 10903 (6.9\%) \\ 
    mean sample dur. (std) [s] & 6.6 ($\pm 5.6$) & 6.6 ($\pm 5.3$) & 6.5 ($\pm 5.4$) & 6.6 ($\pm 5.5$) \\
    total dur. [s] & 13:09:29 & 3:25:55 & 3:21:56 & 19:57:22 \\
    \bottomrule
  \end{tabular}}
\end{table}

\section{Methods}
With both audio samples and their transcripts at our disposal, we conduct three types of experiments. First, we train text-based classifiers, for which we not only use the manual gold standard transcripts but also explore the outputs of various ASR systems (\Cref{ssec:text_methods}). Second, we aim to predict flattery based on the speech samples only, utilizing a range of pretrained audio foundation models (\Cref{ssec:audio_methods}). Third, we seek to combine the merits of text-based and audio-based approaches in one model (\Cref{ssec:multi_methods}). Considering the class imbalance (cf.~\Cref{tab:stats}), we choose \ac{UAR}, 
also known as balanced accuracy as our evaluation metric in all experiments. All models are taken from the 
\texttt{huggingface} hub.

\subsection{Text-based Classification}\label{ssec:text_methods}
We build a text classifier utilizing a pretrained RoBERTa~\cite{liu2019roberta} model in its \emph{base} variant, \ie a $12$-layer Transformer encoder with about $110$M parameters.
We add a classification head after the final layer's encoding and fine-tune all weights of the model utilizing the training partition. The training process runs for at most $7$ epochs but is aborted earlier if no improvement on the development set is observed for two epochs. The learning rate is set to $10^{-5}$ after initial experiments with the values $\{10^{-4}, 5\times10^{-5}, 10^{-5}, 5\times10^{-6}\}$. As the loss function, binary cross-entropy with positive samples weighted inversely to their frequency is utilized. We repeat the training process five times with different fixed random seeds.

Since in practice manual ``gold standard'' transcripts are typically not available, we also explore automatically generated transcripts obtained from different \ac{ASR} models. We consider $6$ different pretrained models from the Whisper~\cite{Whisper} family, ranging from the \emph{tiny} variant with $39$M parameters to \emph{large} models comprising $1.5$B parameters. We generate automatic transcripts using the models without any further adaptation. 
For each of the $6$ \ac{ASR} systems' outputs, we train instances of the RoBERTa classifier described above, applying the same procedure and hyperparameters as for the ``gold standard'' texts. Moreover, we compute the \ac{WER} on the dataset for every \ac{ASR} system. 

\subsection{Audio-based Classification}\label{ssec:audio_methods}
We consider three different types of pretrained audio Transformers, namely variants of \ac{AST}~\cite{gong21b_interspeech}, \ac{W2V}~\cite{baevski2020wav2vec}, and Whisper~\cite{Whisper}.
\ac{AST}~\cite{gong21b_interspeech} is a Transformer model with $12$ layers that takes spectrograms as input.
Specifically, we utilize \ac{AST} trained on the Speech Commands V2 dataset~\cite{warden2018speech},
 as this is the only speech-related model provided in~\cite{gong21b_interspeech}.
\ac{W2V}~\cite{baevski2020wav2vec} 
is pretrained for reconstructing masked parts of speech signals. We employ both the \emph{base} 
($12$ Transformer layers) and \emph{large} 
($24$ Transformer layers) variant of \ac{W2V} which were both pretrained and, subsequently, finetuned on the Librispeech~\cite{panayotov2015librispeech} dataset containing $960$ hours of speech. Moreover, as the task of \ac{SER} is arguably related to our problem of flattery detection, we experiment with a \ac{W2V} model finetuned on MSP-Podcast~\cite{lotfian2017building} for \ac{SER}~\cite{wagner2023dawn}, denoted as \ac{W2V}-MSP.
As for Whisper~\cite{Whisper}, we make use of the \emph{base}, \emph{medium}, and \emph{large} pretrained models.

Our choice of model families is motivated by the finding of~\cite{wagner2023dawn} that models in the fashion of \ac{W2V} acquire linguistic information when fine-tuned. Hence, we opt for a selection of \ac{W2V} models fine-tuned for both \ac{ASR} and \ac{SER}, variants of Whisper,
as a more recent \ac{ASR}-based approach and \ac{AST} that is solely trained on spectrograms and should thus not be equipped with linguistic knowledge. This, in combination with the text-based classification, allows us to reason whether flattery can mainly be recognized via prosody, text or both. 

\subsubsection{Layer-Wise Encodings}\label{sssec:audio_svms}
It has been shown that different layers of pretrained Transformer models for speech encode different acoustic and linguistic properties of an input signal~\cite{pasad2021layer}. Hence, for each model mentioned above, we investigate the aptitude of each of its layers for the flattery detection task. 

First, for every model, we extract representations of the speech signal from each layer. For \ac{AST}, we take the layer's embedding of the special \verb|[CLS]| token as the representation. For Whisper and \ac{W2V} models, representations are obtained by averaging over all of the respective layer's token representations.
We then determine the most promising layer per model by training linear binary classifiers on each layer's representations. Given the large amount of trials, we choose \ac{SVM} classifiers as a lightweight and deterministic option here. We optimize the regularization constant $C$ and the weighting of minority class examples in every experiment.
In the second row of experiments, for every model, we only consider the layer with the best results in the first step and, in addition, the final layer. For both layers' representations, more extensive \ac{SVM} hyperparameter searches are conducted that also optimize the kernel type (RBF, linear, Sigmoid, polynomial), the kernel coefficient $\gamma$, and, if applicable, the degree of the kernel function.

\subsubsection{Audio Model Tuning}\label{sssec:audio_ft}

From each model family (\ac{AST}, \ac{W2V}, Whisper), we first select the variant performing best in the initial SVM experiments (cf.~\Cref{tab:results_audio}) and then finetune the pretrained model by training (i) the full model, and (ii) a version pruned to the layer that performed best among all layers in the SVM experiments.
To do so, we add a linear classification head on top of each pretrained model, similar to the fine-tuning of RoBERTa (cf.~\Cref{ssec:text_methods}).
Analogously to the feature extraction process, we feed the final layers' \verb|[CLS]| encoding for \ac{AST} and the mean over the final layers' token representations for \ac{W2V} and Whisper into the classification head. 
We determine a suitable learning rate for each model by training its pruned version for one epoch with different learning rates ($\{10^{-3}, 10^{-4}, 10^{-5}, 10^{-6}\}$) and three fixed random seeds. Binary Cross Entropy is employed as the loss function. We apply random oversampling to tackle the imbalanced label distribution here. Experiments with a weighted loss function did not yield promising results.
The final models are trained with five different fixed random seeds.

\subsection{Text + Audio Fusion}\label{ssec:multi_methods}
The fusion of the speech and text modality makes use of the models trained on speech only and text only, respectively. We consider the text-based models trained on i) the gold standard transcripts, ii) the weakest \ac{ASR} system's (Whisper-tiny) outputs, and iii) the best \ac{ASR} system's (Whisper-large) outputs (cf.~\Cref{tab:results_text}). For the sake of uniformity, we utilize the best fine-tuned audio model, \ie \ac{W2V}-MSP (cf.~\Cref{tab:results_audio}) for the speech modality. We apply a weighted \textbf{late fusion} on the respective predictions, where the weights for both models are chosen according to their respective performance on the development set. Furthermore, we experiment with \textbf{early fusion}. Specifically, for each pair of audio and text models, we extract their final layers' representations and concatenate them. Then, \ac{SVM} classifiers are trained on these features, analogously to the process described in~\Cref{sssec:audio_svms}. Both the late and early fusion methods are deterministic, however, the models to be fused are all trained for the same five fixed random seeds. Thus, we can report means and standard deviations across these seeds by always fusing the models trained with the same seed.

\section{Results}
In the following, we report the results of the text-based (\Cref{ssec:text_results}), speech-based (\Cref{ssec:audio_results}) and the fusion of text and speech (\Cref{ssec:multi_results}). 

\subsection{Text-based Classification Results}\label{ssec:text_results}
\Cref{tab:results_text} presents the results of training the RoBERTa classifier on different transcripts. In addition, the \acp{WER} of the different \ac{ASR} models are given. 

\begin{table}[th]
  \caption{ASR + Text pipeline results. We report mean \ac{UAR} values and standard deviations across five fixed seeds. \emph{Gold standard} refers to the transcriptions generated by humans.}
  \label{tab:results_text}
  \centering\resizebox{1\columnwidth}{!}{
  \begin{tabular}{lrrrr}
    \toprule
    \multicolumn{1}{l}{Transcriptions} & \multicolumn{1}{l}{\#\,params} & \multicolumn{1}{l}{\%\,WER} & \multicolumn{2}{c}{RoBERTa [\%\,UAR]}\\
     & \multicolumn{1}{r}{(ASR)} & & \multicolumn{1}{c}{dev} & \multicolumn{1}{c}{test} \\
    \midrule
    Whisper-tiny & 39M & 26.60 & 78.79 ($\pm$1.05) & 80.96  ($\pm$0.98)  \\
    Whisper-base & 74M & 20.90 & 81.15 ($\pm$1.44) & 80.23  ($\pm$1.41)  \\
    Whisper-small & 244M & 16.43 & 80.51 ($\pm$2.05) & 83.49  ($\pm$1.14)  \\
    Whisper-medium & 769M & 14.94 & 81.26 ($\pm$1.39) & 83.47  ($\pm$1.35)  \\
    Whisper-large & 1.5B & 14.68 & \textbf{81.68} ($\pm$1.88) & 83.71  ($\pm$1.68)  \\
    Whisper-large-v2 & 1.5B & 14.80 & 79.50 ($\pm$1.65) & 82.71  ($\pm$1.77)  \\
    \midrule
    \textit{gold standard} & - & - & \textbf{82.67} ($\pm$1.69) & 85.97  ($\pm$1.94)  \\
    \bottomrule
  \end{tabular}}
\end{table}

It can be observed that the larger Whisper \ac{ASR} models perform better than the smaller ones regarding their \ac{WER}, with the \emph{tiny} model producing texts with a \ac{WER} of $26.60$\,\% while the \acp{WER} of the \emph{medium}, and \emph{large} variants are around 

$15.00$\,\%. Regarding the flattery classification, the best average result of $82.67\,\%$ \ac{UAR} on the development and $85.97\,\%$ \ac{UAR} on the test set is achieved when training with the gold standard transcripts. All results prove to be stable across seeds, as no standard deviation exceeds $2\,\%$ on the test set. While the gold standard transcriptions model outperforms the best ASR transcriptions model, \ie \emph{Whisper-large}, by more than $2$ percentage points on the test data, all \ac{ASR} transcript-based models still achieve over $80\,\%$ mean \ac{UAR} on the test set. This indicates that for the task of textual flattery detection, even relatively high \acp{WER} such as $26.60$\,\% for \emph{Whisper-tiny} are not too detrimental to the text classifier's performance. One explanation for this is that high \acp{WER} in this particular data set are mainly due to highly domain-specific terms that carry no information related to flattery and are thus less relevant for the classification. Nevertheless, there is a connection between \acp{WER} and the corresponding classification results, with \emph{Whisper-tiny} being responsible for the worst result on the development set ($78.79\,\%$ UAR) while the best \ac{ASR}-based classification result on the development set ($81.68\,\%$ \ac{UAR}) is achieved with the transcripts of \emph{Whisper-large} that have the lowest \ac{WER} among all the \ac{ASR} models ($14.68$\,\%).

\subsection{Audio-based Classification Results}\label{ssec:audio_results}
The results for the audio-based flattery detection with both \acp{SVM} and finetuning are given in~\Cref{tab:results_audio}.

\begin{table}[th]
  \caption{Results of the audio-based experiments. For the finetuning experiments, mean \ac{UAR} values and standard deviations across five fixed seeds are given. The best \ac{SVM} result per model family on the development set is underlined; the best development results overall are boldfaced for both SVMs and fine-tuned models.}
  \label{tab:results_audio}
  \centering
  \resizebox{\columnwidth}{!}{
  \begin{tabular}{llrrrr}
    \toprule
    \multicolumn{1}{l}{Model} & \multicolumn{1}{l}{Layer} & \multicolumn{2}{c}{SVM [UAR]} & \multicolumn{2}{c}{Finetuning [UAR]} \\
     & & \multicolumn{1}{c}{dev} & \multicolumn{1}{c}{test} & \multicolumn{1}{c}{dev} & \multicolumn{1}{c}{test} \\
    \midrule
    
    \ac{AST} & 4 & \underline{57.49} & 51.34 & 56.32 ($\pm$.1.46) & 51.99  ($\pm$1.70) \\
    \ac{AST} & 12 & 55.85 & 54.46 & 52.41 ($\pm$.0.60) & 53.44  ($\pm$0.42)  \\
    \midrule
    
    W2V-base & 7 & 75.36 & 72.94 & - & -  \\
    W2V-base & 12 & 66.84 & 62.63 & - & -  \\
    
    W2V-large & 11 & 78.45 & 75.60 & - & -  \\
    W2V-large & 24 & 73.70 & 69.17 & - & -  \\
    
    W2V-MSP & 11 & 79.70 & 82.23 & - & -  \\
    W2V-MSP & 12 & \textbf{\underline{79.71}} & 82.46 & \textbf{78.94} ($\pm$0.64) & 80.60  ($\pm$0.58)  \\
    \midrule
        
    Whisper-base & 5 & 69.27 & .69.13 & - & -  \\
    Whisper-base & 6 & 70.04 & 66.62 & - & -  \\
    
    Whisper-medium & 23 & \underline{79.46} & 76.31 & 72.32 ($\pm$6.44) & 74.52 ($\pm$6.35)  \\
    Whisper-medium & 24 & 79.37 & 75.52 & 76.94 ($\pm$2.83) & 78.91 ($\pm$2.26) \\
    
    Whisper-large & 29 & 78.54 & 72.61 & - & -  \\
    Whisper-large & 32 & 77.05 & 76.28 & - & -  \\
    \bottomrule
  \end{tabular}}
\end{table}

The \ac{AST} experiments yield considerably worse results than those based on the different \ac{W2V} and Whisper variants. While most results of \ac{W2V} and Whisper exceed $70\,\%$ \ac{UAR}, all \ac{AST}-based experiments only slightly surpass the chance level of $50\,\%$ \ac{UAR}. Considering the \ac{UAR} values of over $80\,\%$ observed in the text-based experiments, we assume that this performance gap is partially due to \ac{W2V} and Whisper encoding linguistic information, which is not the case for \ac{AST}. Consequently, the low \ac{UAR} values for \ac{AST} suggest that flattery can rarely be detected via prosodic information only. Another aspect that may contribute to \ac{AST}'s rather poor performance is that the SpeechCommand data it is initially trained on differs from our data in that all its speech samples are only one second long. Lastly, as our data is obtained from calls, the audio quality may be impaired, suppressing prosodic attributes of the speech samples that might prove beneficial for audio-based classification.

The layer-wise results confirm that different layers of pretrained models are of different suitability for the flattery detection task. This is particularly prominent for the \ac{W2V} variants, where layer $7$ clearly outperforms the final layer ($12$) in the base model and layer $11$ leads to a considerably better result ($75.60\,\%$) on the test set than layer $24$ ($62.63\,\%$) in the large model. As for the Whisper models, the best layers are always close, but never identical, to the ultimate layer. All \ac{W2V} and Whisper variants yield results better than $75\,\%$ \ac{UAR} on the development set in their best layer in the SVM results, with \ac{W2V}-MSP achieving the best \ac{UAR} values overall on both the development ($79.71\,\%$) and test ($82.46\,\%$) set. 

Finetuning generally does not improve upon the SVM results. The standard deviation of $6.44$ for Whisper-medium, however, shows that, depending on the random seed, results over $80\,\%$ \ac{UAR} on the test set are possible. Overall, the best audio-based classifiers perform slightly worse than the best text-based classifiers that achieve over $83\,\%$ mean \ac{UAR} on the test set, cf.~\Cref{tab:results_text}.


\subsection{Text + Audio Fusion Results}\label{ssec:multi_results}
We report the multimodal results in~\Cref{tab:results_multi}.

\begin{table}[th]

\caption{Results for the experiments fusing audio (A) and textual (T) information. We provide means and standard deviations across five fixed seeds, where the best result on development per transcription method is underlined, while the best overall is boldfaced. \emph{T only} refers to the textual experiments reported in~\Cref{tab:results_text} for reference.}\label{tab:results_multi}
  \centering
  \resizebox{\columnwidth}{!}{
  \begin{tabular}{llrr}
    \toprule
    Transcriptions & Method & \multicolumn{2}{c}{[UAR]} \\
     & & \multicolumn{1}{c}{dev} & \multicolumn{1}{c}{test} \\
    \midrule
    
    Whisper-tiny & T only & 78.79 ($\pm$1.05) & 80.96 ($\pm$0.98) \\
     & Late Fusion A+T & 79.72 ($\pm$1.50) & 82.12 ($\pm$1.70) \\
     & Early Fusion A+T & \underline{81.85} ($\pm$2.04) & 83.69 ($\pm$1.86) \\
     \midrule

     Whisper-large & T only & 81.68 ($\pm$1.88) & 83.71 ($\pm$1.68) \\
     & Late Fusion A+T & 82.02 ($\pm$1.90) & 83.94 ($\pm$1.39) \\
     & Early Fusion A+T & \underline{83.62} ($\pm$1.56) & 84.71 ($\pm$1.01) \\
     \midrule

     \textit{gold standard} & T only & 82.67 ($\pm$1.69) & 85.97 ($\pm$1.94) \\
     & Late Fusion A+T & 83.02 ($\pm$1.56) & 86.41 ($\pm$1.86) \\
     & Early Fusion A+T & \underline{\textbf{84.80}} ($\pm$1.33) & 87.16 ($\pm$1.33) \\
    \bottomrule
  \end{tabular}}
\end{table}

It is evident that for all transcripts considered, a combination with the speech modality improves upon the text-only approach. Hence, it can be assumed that our speech-based models, though arguably also making use of linguistic information, encode information that complements the text-only representations to a degree. A comparison of late and early fusion shows that in all cases, the early fusion approach outperforms the late fusion method. The comparably weak performance of the latter may be attributed to the poor calibration we observe in our fine-tuned Transformers' predictions.

Among the different transcripts, the largest improvement over the purely textual model can be observed for those generated with Whisper-tiny, \ie the worst performing \ac{ASR} system (cf.~\Cref{tab:results_text}). Specifically, its mean early fusion \ac{UAR} value on the development set ($81.85\,\%$) exceeds its mean text-only \ac{UAR} result ($78.79\,\%$) by $3.88\,\%$. The relative improvement is lower for the transcripts obtained via Whisper-large and the gold standard, namely $2.38\,\%$ and $2.58\,\%$, respectively. This suggests that the audio modality can complement text-only approaches to flattery detection, especially when the \ac{ASR} system's \ac{WER} is relatively high. A closer manual inspection of data points for which audio and text classifiers disagree reveals another class of instances that benefit from speech-based classification: certain phrases, such as variants of \textit{Great!} and \textit{Good morning}, are sometimes labeled as flattery and sometimes not, depending on their context. Thus, as our simple text-based models do not consider the surrounding sentences, audio-based classifiers prove helpful in correctly predicting flattery in such utterances.   

\subsection{Generalization to Female Speakers}
Given that women are considerably underrepresented in our dataset (cf.~\Cref{tab:stats}), we investigate our models' generalizability for female speakers. In \Cref{tab:results_mf}, the results of the best finetuned Transformer models are broken down into female and male speakers. While for both the text and the audio approach, the \ac{UAR} values for women are lower than those for men, they are typically close. The largest gap is observed for the text-based predictions on the development set, with the \ac{UAR} for females ($75.99$) being about $7$ percentage points lower than that for males ($83.27$) on average. It is also evident that the results for the comparatively few female data points tend to vary more depending on the random seed. An explanation for the rather small gap in performance for female and male speakers may be that, at least in the business context, there may not be many gender-based differences when it comes to using flattering phrases. As the \ac{W2V} models arguably also draw heavily on linguistic information, this reasoning would apply to them as well. 

\section{Discussion}

We observe that the textual modality, \ie \emph{what} is said, is crucial for predicting flattery. Second, the speech signal, while yielding less promising results on its own, still encodes valuable information that complements and thus improves text-based classification -- especially in cases where the automatic transcription of utterances performs comparably poorly. Besides, the speech-based experiments with \ac{AST}, Whisper, and \ac{W2V} again demonstrate that fine-tuned \ac{ASR}-based audio foundation models encode both linguistic and prosodic information. 

Potential limitations to the generalisability of our models are induced by the nature of the data set. As the data is sourced from business analyst calls in US companies, it is arguably highly context-specific and not representative of the general population with respect to demographic aspects such as educational background or age. 

\begin{table}
\caption{Results of RoBERTa finetuned on the gold standard transcripts (cf.~\Cref{tab:results_text}) and finetuned \ac{W2V}-MSP (cf.~\Cref{tab:results_audio}) for female (F) and, respectively, male (M) speakers only. We report means and standard deviations across $5$ fixed seeds.}
\label{tab:results_mf}
  \centering
  \resizebox{\columnwidth}{!}{
  \begin{tabular}{llrr}
    \toprule
    Approach & Subset & \multicolumn{2}{c}{[UAR]} \\
     & & \multicolumn{1}{c}{dev} & \multicolumn{1}{c}{test} \\
    \midrule
    
    RoBERTa & F only & 75.99 ($\pm$4.75) & 83.76 ($\pm$4.68) \\
     & M only & 83.27 ($\pm$1.47) & 86.21 ($\pm$1.91) \\
    
     \midrule

     W2V-MSP (finetuned) & F only & 77.99 ($\pm$3.59) & 84.21 ($\pm$2.59) \\
     & M only & 78.91 ($\pm$0.65) & 80.21 ($\pm$0.83) \\

    \bottomrule
  \end{tabular}}
\end{table}

\section{Conclusion}

We introduced the problem of flattery detection from speech alongside a novel data set. Furthermore, we trained an extensive set of \ac{ML} approaches based on speech, text, and the combination of both modalities, thus providing insights into the nature of this novel task. 

Future work may include extending the database to cover broader demographics. Regarding the methodology, considering larger textual units in order to capture the sentences' contexts better is a promising avenue. Moreover, more refined fusion methods than those utilized in~\Cref{ssec:multi_methods} can be devised. 
While we cannot publish our raw data due to copyright restrictions, we make our code, extracted features, and the best-performing models available\footnote{\url{https://github.com/lc0197/flattery\_from\_speech}}.

\section{Acknowledgements}
This work was supported by MDSI -- Munich Data Science Institute as well as MCML -- Munich Center of Machine Learning.
Bj\"orn W. Schuller is also with the Konrad Zuse School of Excellence in Reliable AI (relAI), Munich, Germany. 
\newpage
\bibliographystyle{IEEEtran}
\bibliography{mybib}

\begin{thebibliography}{10}
\providecommand{\url}[1]{#1}
\csname url@samestyle\endcsname
\providecommand{\newblock}{\relax}
\providecommand{\bibinfo}[2]{#2}
\providecommand{\BIBentrySTDinterwordspacing}{\spaceskip=0pt\relax}
\providecommand{\BIBentryALTinterwordstretchfactor}{4}
\providecommand{\BIBentryALTinterwordspacing}{\spaceskip=\fontdimen2\font plus
\BIBentryALTinterwordstretchfactor\fontdimen3\font minus \fontdimen4\font\relax}
\providecommand{\BIBforeignlanguage}[2]{{%
\expandafter\ifx\csname l@#1\endcsname\relax
\typeout{** WARNING: IEEEtran.bst: No hyphenation pattern has been}%
\typeout{** loaded for the language `#1'. Using the pattern for}%
\typeout{** the default language instead.}%
\else
\language=\csname l@#1\endcsname
\fi
#2}}
\providecommand{\BIBdecl}{\relax}
\BIBdecl

\bibitem{jones1975ingratiation}
E.~E. Jones, \emph{Ingratiation: A social psychological analysis}.\hskip 1em plus 0.5em minus 0.4em\relax S.I.: Irvington Publishers, 1975.

\bibitem{kumar1991construction}
K.~Kumar and M.~Beyerlein, ``Construction and validation of an instrument for measuring ingratiatory behaviors in organizational settings.'' \emph{Journal of applied psychology}, vol.~76, no.~5, pp. 619--627, 1991.

\bibitem{westphal1998board}
J.~D. Westphal, ``Board games: How ceos adapt to increases in structural board independence from management,'' \emph{Administrative science quarterly}, pp. 511--537, 1998.

\bibitem{vonk2007ingratiation}
R.~Vonk, ``Ingratiation,'' in \emph{Encyclopedia of social psychology}, R.~F. Baumeister, Ed.\hskip 1em plus 0.5em minus 0.4em\relax Sage, 2007, pp. 481--483.

\bibitem{gordon1996impact}
R.~A. Gordon, ``Impact of ingratiation on judgments and evaluations: A meta-analytic investigation.'' \emph{Journal of personality and social psychology}, vol.~71, no.~1, p.~54, 1996.

\bibitem{ellis2002use}
A.~P. Ellis, B.~J. West, A.~M. Ryan, and R.~P. DeShon, ``The use of impression management tactics in structured interviews: A function of question type?'' \emph{Journal of applied psychology}, vol.~87, no.~6, p. 1200, 2002.

\bibitem{westphal2008sociopolitical}
J.~D. Westphal and M.~B. Clement, ``Sociopolitical dynamics in relations between top managers and security analysts: Favor rendering, reciprocity, and analyst stock recommendations,'' \emph{Academy of Management Journal}, vol.~51, no.~5, pp. 873--897, 2008.

\bibitem{westphal2011avoiding}
J.~D. Westphal and D.~L. Deephouse, ``Avoiding bad press: Interpersonal influence in relations between ceos and journalists and the consequences for press reporting about firms and their leadership,'' \emph{Organization Science}, vol.~22, no.~4, pp. 1061--1086, 2011.

\bibitem{mcewen1997communication}
T.~McEwen, ``Communication training in corporate settings: Lessons and opportunities for the academe,'' \emph{American Journal of Business}, vol.~12, no.~1, pp. 49--58, 1997.

\bibitem{cipresso2017back}
P.~Cipresso and J.~C. Immekus, ``Back to the future of quantitative psychology and measurement: psychometrics in the twenty-first century,'' p. 2099, 2017.

\bibitem{gerczuk2021emonet}
M.~Gerczuk, S.~Amiriparian, S.~Ottl, and B.~W. Schuller, ``Emonet: a transfer learning framework for multi-corpus speech emotion recognition,'' \emph{IEEE Transactions on Affective Computing}, 2021.

\bibitem{pepino21_interspeech}
L.~Pepino, P.~Riera, and L.~Ferrer, ``{Emotion Recognition from Speech Using wav2vec 2.0 Embeddings},'' in \emph{Proc. INTERSPEECH}, 2021, pp. 3400--3404.

\bibitem{wagner2023dawn}
J.~Wagner, A.~Triantafyllopoulos, H.~Wierstorf, M.~Schmitt, F.~Burkhardt, F.~Eyben, and B.~W. Schuller, ``Dawn of the transformer era in speech emotion recognition: Closing the valence gap,'' \emph{IEEE Transactions on Pattern Analysis \& Machine Intelligence}, vol.~45, no.~09, pp. 10\,745--10\,759, 2023.

\bibitem{christ2023towards}
L.~Christ, S.~Amiriparian, A.~Kathan, N.~M{\"u}ller, A.~K{\"o}nig, and B.~W. Schuller, ``Towards multimodal prediction of spontaneous humour: A novel dataset and first results,'' \emph{arXiv preprint arXiv:2209.14272}, 2023.

\bibitem{amiriparian2022muse}
S.~Amiriparian, L.~Christ, A.~K{\"o}nig, E.-M. Me{\ss}ner, A.~Cowen, E.~Cambria, and B.~W. Schuller, ``Muse 2022 challenge: Multimodal humour, emotional reactions, and stress,'' in \emph{Proc. ACM Multimedia}, 2022, pp. 7389--7391.

\bibitem{hasan2019ur}
M.~K. Hasan, W.~Rahman, A.~Bagher~Zadeh, J.~Zhong, M.~I. Tanveer, L.-P. Morency, and M.~E. Hoque, ``{UR}-{FUNNY}: A multimodal language dataset for understanding humor,'' in \emph{Proc. EMNLP-IJCNLP}.\hskip 1em plus 0.5em minus 0.4em\relax Hong Kong, China: Association for Computational Linguistics, Nov. 2019, pp. 2046--2056.

\bibitem{bedi2021multi}
M.~Bedi, S.~Kumar, M.~S. Akhtar, and T.~Chakraborty, ``Multi-modal sarcasm detection and humor classification in code-mixed conversations,'' \emph{IEEE Transactions on Affective Computing}, 2021.

\bibitem{amiriparian2019synchronization}
S.~Amiriparian, J.~Han, M.~Schmitt, A.~Baird, A.~Mallol-Ragolta, M.~Milling, M.~Gerczuk, and B.~Schuller, ``Synchronization in interpersonal speech,'' \emph{Frontiers in Robotics and AI}, vol.~6, p. 116, 2019.

\bibitem{defensive2023}
S.~Amiriparian, L.~Christ, R.~Kushtanova, M.~Gerczuk, A.~Teynor, and B.~W. Schuller, ``Speech-based classification of defensive communication: A novel dataset and results,'' in \emph{Proc. INTERSPEECH}, 2023, pp. 2703 -- 2707.

\bibitem{amiriparian2021impact}
S.~Amiriparian, A.~Sokolov, I.~Aslan, L.~Christ, M.~Gerczuk, T.~H{\"u}bner, D.~Lamanov, M.~Milling, S.~Ottl, I.~Poduremennykh \emph{et~al.}, ``On the impact of word error rate on acoustic-linguistic speech emotion recognition: An update for the deep learning era,'' \emph{arXiv preprint arXiv:2104.10121}, 2021.

\bibitem{gandhi2023multimodal}
A.~Gandhi, K.~Adhvaryu, S.~Poria, E.~Cambria, and A.~Hussain, ``Multimodal sentiment analysis: A systematic review of history, datasets, multimodal fusion methods, applications, challenges and future directions,'' \emph{Information Fusion}, vol.~91, pp. 424--444, 2023.

\bibitem{christ2023muse}
L.~Christ, S.~Amiriparian, A.~Baird, A.~Kathan, N.~M{\"u}ller, S.~Klug, C.~Gagne, P.~Tzirakis, L.~Stappen, E.-M. Me{\ss}ner \emph{et~al.}, ``The muse 2023 multimodal sentiment analysis challenge: Mimicked emotions, cross-cultural humour, and personalisation,'' in \emph{Proc. MuSe}, 2023, pp. 1--10.

\bibitem{konig2018silver}
A.~K{\"o}nig, J.~Mammen, J.~Luger, A.~Fehn, and A.~Enders, ``Silver bullet or ricochet? ceos’ use of metaphorical communication and infomediaries’ evaluations,'' \emph{Academy of Management Journal}, vol.~61, no.~4, pp. 1196--1230, 2018.

\bibitem{krippendorff2018content}
K.~Krippendorff, \emph{Content analysis: An introduction to its methodology}.\hskip 1em plus 0.5em minus 0.4em\relax Sage publications, 2018.

\bibitem{schill2022analyst}
A.-K. Schill, A.~Boutalikakis, F.~Hawighorst, L.~Graf-Vlachy, and A.~S. Konig, ``Analyst flattery, ceo narcissism, and ceo communication specificity,'' in \emph{Academy of Management Proceedings}, vol. 2022, no.~1.\hskip 1em plus 0.5em minus 0.4em\relax Academy of Management Briarcliff Manor, NY 10510, 2022, p. 10892.

\bibitem{sadvilkar-neumann-2020-pysbd}
N.~Sadvilkar and M.~Neumann, ``{P}y{SBD}: Pragmatic sentence boundary disambiguation,'' in \emph{Proc. NLP-OSS}, E.~L. Park, M.~Hagiwara, D.~Milajevs, N.~F. Liu, G.~Chauhan, and L.~Tan, Eds.\hskip 1em plus 0.5em minus 0.4em\relax Online: Association for Computational Linguistics, Nov. 2020, pp. 110--114.

\bibitem{mcauliffe2017montreal}
M.~McAuliffe, M.~Socolof, S.~Mihuc, M.~Wagner, and M.~Sonderegger, ``Montreal forced aligner: Trainable text-speech alignment using kaldi.'' in \emph{Proc. INTERSPEECH}, vol. 2017.\hskip 1em plus 0.5em minus 0.4em\relax Stockholm, Sweden: International Speech Communication Association (ISCA), 2017, pp. 498--502.

\bibitem{liu2019roberta}
Y.~Liu, M.~Ott, N.~Goyal, J.~Du, M.~Joshi, D.~Chen, O.~Levy, M.~Lewis, L.~Zettlemoyer, and V.~Stoyanov, ``Roberta: A robustly optimized bert pretraining approach,'' \emph{arXiv preprint arXiv:1907.11692}, 2019.

\bibitem{Whisper}
A.~Radford, J.~W. Kim, T.~Xu, G.~Brockman, C.~McLeavey, and I.~Sutskever, ``Robust speech recognition via large-scale weak supervision,'' in \emph{Proceedings of the 40th International Conference on Machine Learning}, ser. ICML'23.\hskip 1em plus 0.5em minus 0.4em\relax JMLR.org, 2023.

\bibitem{gong21b_interspeech}
Y.~Gong, Y.-A. Chung, and J.~Glass, ``{AST: Audio Spectrogram Transformer},'' in \emph{Proc. INTERSPEECH}, 2021, pp. 571--575.

\bibitem{baevski2020wav2vec}
A.~Baevski, Y.~Zhou, A.~Mohamed, and M.~Auli, ``wav2vec 2.0: A framework for self-supervised learning of speech representations,'' \emph{Proc. NeurIPS}, vol.~33, pp. 12\,449--12\,460, 2020.

\bibitem{warden2018speech}
P.~Warden, ``Speech commands: A dataset for limited-vocabulary speech recognition,'' \emph{arXiv preprint arXiv:1804.03209}, 2018.

\bibitem{panayotov2015librispeech}
V.~Panayotov, G.~Chen, D.~Povey, and S.~Khudanpur, ``Librispeech: an asr corpus based on public domain audio books,'' in \emph{Proc. ICASSP}.\hskip 1em plus 0.5em minus 0.4em\relax IEEE, 2015, pp. 5206--5210.

\bibitem{lotfian2017building}
R.~Lotfian and C.~Busso, ``Building naturalistic emotionally balanced speech corpus by retrieving emotional speech from existing podcast recordings,'' \emph{IEEE Transactions on Affective Computing}, vol.~10, no.~4, pp. 471--483, 2017.

\bibitem{pasad2021layer}
A.~Pasad, J.-C. Chou, and K.~Livescu, ``Layer-wise analysis of a self-supervised speech representation model,'' in \emph{2021 IEEE Automatic Speech Recognition and Understanding Workshop (ASRU)}.\hskip 1em plus 0.5em minus 0.4em\relax IEEE, 2021, pp. 914--921.

\end{thebibliography}
\end{document}